# Atomic structure and magnetism of the Au-Ga-Ce 1/1 approximant crystal

Shintaro Suzuki[1,*,†], Azusa Motouri[1*], Kazuhiko Deguchi[2], Tsunetomo Yamada[3], Asuka Ishikawa[4], Takenori Fujii[5], Kazuhiro Nawa[6], Taku J. Sato[6], and Ryuji Tamura[1]

[1] *Department of Material Science and Technology, Tokyo University of Science, Niijuku, Tokyo 125-8585, Japan*
[2] *Department of Physics, Graduate School of Science, Nagoya University, Nagoya 464-8602, Japan*
[3] *Department of Applied Physics, Tokyo University of Science, Niijuku, Tokyo 125–8585, Japan*
[4] *Research Institute for Science and Technology, Tokyo University of Science, Niijuku, Tokyo 125–8585, Japan*
[5] *Cryogenic Research Center, The University of Tokyo, Bunkyo, Tokyo 113-0032, Japan*
[6] *Institute of Multidisciplinary Research for Advanced Materials, Tohoku University, Sendai 980-8577, Japan*

We report a new Au-Ga-Ce 1/1 approximant crystal (AC) which possesses a significantly wide single-phase region of 53 - 70 at% Au and 13.6 - 15.1 at% Ce. Single crystal X-ray structural analyses reveal the existence of two types of structural degrees of freedom, i.e., the Au/Ga mixing sites and the fractional Ce occupancy site: the former enables a large variation in the electron concentration and the latter allows a variation in the occupancy of a magnetic impurity atom at the center of the Tsai-type cluster. Following these findings, the influences of two types of structural modifications on the magnetism are thoroughly investigated by means of magnetic susceptibility and specific heat measurements on the Au-Ga-Ce 1/1 AC. The spin-glass (SG) state is found to be the ground state over the entire single-phase region, showing a robust nature of the SG state against both structural modifications. In addition, a gigantic specific heat ($C/T$) is commonly observed at low temperatures for all the compositions, which is consistently explained as a consequence of the spin-freezing phenomenon, not of a heavy Fermion behavior as reported elsewhere. Moreover, the origin of the SG state in the 1/1 Au-Ga-Ce AC is attributed to the existence of non-magnetic atom disorder in the Au/Ga mixing sites. Furthermore, a Kondo behavior is observed in the electrical resistivity at low temperatures, which is enhanced by increasing the Ce concentration, verifying that a Ce atom introduced at the cluster center behaves as a Kondo impurity for the first time.

## I. INTRODUCTION

The discovery of a quasicrystal (QC) by Shechtman in 1984 has opened a new research field of quasiperiodicity in condensed matter physics [1]. Especially, the relationship between quasiperiodic lattices and their physical properties has attracted considerable attention [2-6]. Recently, magnetism and strongly correlation behavior in quasiperiodic lattices have been extensively investigated in Tsai-type compounds [7], where the magnetic atoms are located only at specific crystallographic sites [8] in sharp contrast with the other icosahedral QCs such as Mackay-type [9] and the Bergman-type [10]. The Tsai-type cluster is composed of five concentric successive shells as described in Fig. 1(a) [12]; a tetrahedron, a dodecahedron, an icosahedron, an icosidodecahedron and a rhombic triacontahedron, where the orientation of the tetrahedron is disordered at room temperature in most cases [13]. Interestingly, recent work has shown that the tetrahedron can be replaced by a single large atom such as a Lanthanoid in some systems [14].

Tsai-type QCs and approximant crystals (ACs) are composed of the same cluster unit but differ in its arrangement, i.e., quasiperiodic or periodic. In Tsai-type 1/1 ACs, its structure can be described as a bcc array of the Tsai-type cluster as shown in Fig. 1(b). Here, we note that one more building block, i.e., an acute rhombohedron, is necessary for construction of QCs, and higher-order ACs [8,15]. In Tsai-type compounds, magnetic Lanthanoids are located on the vertices of the icosahedron and/or at the center of the Tsai-type cluster.

Recently, the magnetism of Tsai-type QCs and ACs with Lanthernoids having a large magnetic moment such as Gd, Tb, and Eu has been extensively investigated. Observations of ferromagnetism and antiferromagnetism were reported in many 1/1 and 2/1 Tsai-type ACs [16-20] and their magnetic ground states were found to be well classified by the electron-per-atom $e/a$ [21] in Gd and Tb-based ACs. The paramagnetic Curie temperature was also found to be well controllable by $e/a$. Based on the phase diagram of ACs, recently, the first long-range ferromagnetic order was discovered in *i*-Au-Ga-Gd, Tb [22] and *i*-Au-Ga-Dy [23] by tuning the $e/a$ value inside the ferromagnetic region of ACs. Magnetic structures formed by spins located at the icosahedral vertices were also extensively investigated experimentally [24,25] as well as theoretically [26-29]; the theoretical work suggests the possible occurrence of various exotic magnetic structures. The effect of the introduction of a magnetic Lanthanoid at the cluster center on the magnetism was also investigated in the Au-Si-Tb 1/1 AC [14, 30], which showed that the spin-glass state is stabilized by the introduction of magnetic atoms.

--------
\* These authors contributed equally to this work.
† Present address: Department of Physical Sciences, Aoyama Gakuin University, Sagamihara, 252-5258, Japan

…

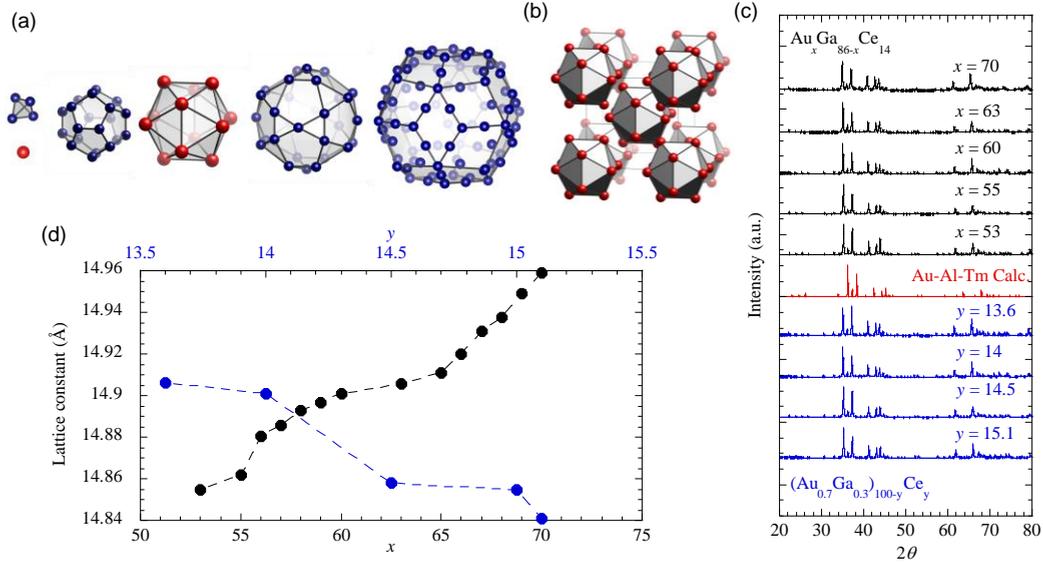

Figure 1(a) The schematic picture of Tsai-type rhombic triacontahedron (RTH) cluster. Magnetic elements locate on vertices of icosahedral shell and the center of the cluster with displacing the tetrahedron shell. (b) Arrangements of icosahedral elements in 1/1 AC. Figure 1(a) and (b) are depicted by VESTA software [11]. (c) The powder X-ray patterns of $Au_xGa_{86-x}Ce_{14}$ (black) and $(Au_{0.70}Ga_{0.30})_{100-y}Ce_y$ (blue), respectively. The calculated diffraction pattern by structural model of Au-Al-Tm Tsai-type 1/1 approximants is also displayed as a reference (red). All observed peaks can be assigned to Tsai-type 1/1 approximants. (d) The $x$ (black) and $y$ (blue) dependence of lattice constant. Increasing Au amount affects as a negative chemical pressure on the system, while increasing Ce amount does as the positive chemical pressure.

Tsai-type QCs and ACs with a small magnetic moment such as Yb have also become a target of growing interest. A non-Fermi liquid behavior was discovered in the Au-Al-Yb QC without tuning, namely, at ambient pressure and at zero field [31-34]. Since the temperature dependence of the physical property near the quantum criticality point is similar to those of $\beta$-$YbAlB_4$ [35,36] and $YbRh_2Si_2$ [37], theoretical works pointed out that the origin of the critical behavior of the QC is due to valence fluctuation [38,39]. Low temperature physical properties were also studied in the Ag-In-Ce [40], Au-Ge-Ce [41] and Au-Al-Ce [42] ACs, in particular, from the viewpoint of the electron-hole symmetry with Yb. Extraordinarily large heat capacity ($C/T$) was commonly observed in all the Ce-based ACs; over 4 J/K$^2$ mol-Ce, the reason of which has not been clarified yet. For the Ag-In-Ce and Au-Ge-Ce ACs, the magnetic ground state is spin-glass whereas Kondo effect is observed for the Au-Al-Ce AC, implying the important role played by the $c$-$f$ hybridization for the latter.

For Ce-based compounds such as heavy fermion compounds [43, 44], introduction of chemical pressure or electron/hole doping via substitution of non-magnetic elements has been frequently performed in order to change the ground state based on the Doniach phase diagram. However, it has not been the case for Ce-based ACs up to date. Although the physical property was investigated at several compositions for the Au-Al-Ce AC, however, its phase was reported to change from 1/1 to 2/1 AC with composition, thus not allowing the tuning of the physical property inside the single phase region. Moreover, the effect of introduction of a rare-earth atom at the cluster center on physical properties has never been reported in Ce based compounds. In this article, we report a new Ce based 1/1 Au-Ga-Ce AC with rich structural degrees of freedom, and discuss the relationship between the physical property and two types of structural modifications. The origin of the SG behavior, a gigantic specific heat and Kondo effect observed in the 1/1 Au-Ga-Ce AC will be addressed based on the structural models obtained by single crystal X-ray structural analyses.

## II. METHOD

Polycrystalline alloys of Au-Ga-Ce were prepared by arc-melting method with using high purity elements of Au (Tanaka Kikinzoku Kogyo K.K.; 99.99%), Ga (Furuuchi Chemical Co.; 99.9999%), Ce (Mitsui Kinzoku Trading; 99.9%). The synthesized alloys were annealed at 873 K at 50h under the Ar atmosphere. The phase purity and the lattice constant were examined by powder X-ray diffraction measurement (Rigaku; MiniFlex600). For crystal structure analysis, intensity data of the diffraction were collected using a RIGAKU Saturn CCD single-crystal diffractometer incorporating VariMax confocal optics for Mo K$\alpha$ radiation ($\lambda$ = 0.71073 Å). Indexing of reflections, integration, and absorption corrections for intensities were executed using the CrysAlisPro software package (Rigaku Oxford Diffraction).

The magnetic susceptibility of the Au-Ga-Ce 1/1 ACs was measured by using commercial magnetic property measurement system (MPMS; Quantum design) with using a SQUID sensor from 2 to 300 K. Low temperature ac susceptibility measurement was also performed by mutual inductance method with a modulation field having an amplitude of 0.1 Oe at a frequency of 301.7 Hz between 0.25 and 5 K. The heat capacity was measured by using physical property measurement system (PPMS; Quantum design) with using relaxation method from 2 to 20 K and by the quasi-adiabatic heat-pulse method from 0.25 to 20 K using a commercial $^3$He refrigerator (Heliox, Oxford Instrument). Measurements of transport properties were performed with 4-probe method from 3 to 300 K.

## III. RESULTS

### A. Samples Characterization

Figure 1(c) shows powder X-ray diffraction patterns of $Au_xGa_{86-x}Ce_{14}$ and $(Au_{0.70}Ga_{0.30})_{100-y}Ce_y$ with various Au concentrations from

…

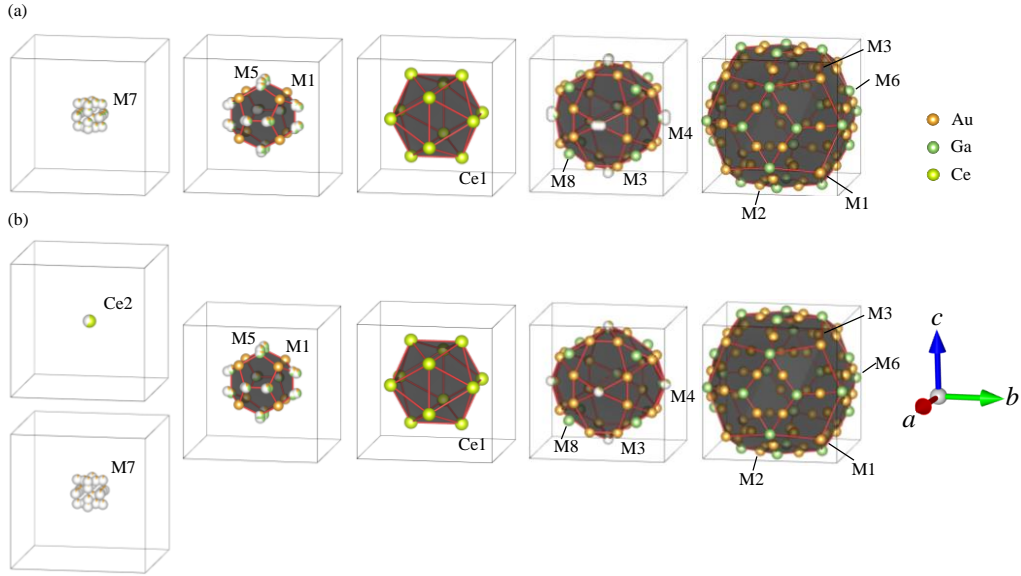

Figure 2 (a), (b) Shell structures of $y = 13.6$ and 15.1 obtained from the single-crystal refinements, respectively. The center of the cluster is fully occupied by a tetrahedron consisted by nonmagnetic Au/Ga elements for $y = 13.6$. On the other hand, 58.1% of the center is replaced by a magnetic Ce element for $y = 15.1$. These shells are depicted by using VESTA software [11].

$x = 53$ to 70 and Ce concentrations from $y = 13.6$ to 15.1, respectively. In the figure, all the observed peaks can be assigned as those of the 1/1 Tsai-type AC, which shows that a single phase is obtained for all the samples indicating that the Au-Ga-Ce 1/1 AC possesses a large single-phase region. Figure 1(d) shows the $x$ and $y$ dependences of the lattice constant. The $x$ dependence can be attributed to the radius difference of Au and Ga, where substitution of Ga for Au may be regarded as introduction of negative chemical pressure. The lattice constant decreases with increasing $y$, which implies the replacement of a tetrahedron by a rare-earth element in the Tsai-type cluster as reported elsewhere [30]. To clarify this point, we then performed structural analyses single-crystal diffraction intensities for the samples with $y = 13.6$ and 15.1.

### B. Structural analyses

The crystallographic information and experimental details are provided in Table 1. For the $y = 13.6$ and 15.1 samples, 48248 and 45550 reflections peaks were observed with resolution limit of approximately 0.5 Å, and the reflections were indexed using a body-centered cubic lattice with a unit-cell parameter of 14.88720(5) and 14.82600(3) Å, respectively. The intensity integration of the reflections was carried out, which was followed by the absorption correction. For both samples, the space group $Im\bar{3}$ was suggested by the GRAL module in the CrysAlisPro, which led to 2631 and 2446 independent reflections with internal agreement factor ($R_{int}$) being 0.0652 and 0.0871 for the $y = 13.6$ and 15.1 samples, respectively. The unit-cell parameter of the $y = 13.6$ sample was larger than that of the $y = 15.1$ sample by approximately 0.061 Å, which is in good agreement with the result by powder diffraction where the unit-cell parameters was observed to increase by approximately 0.065 Å as the $y$ decreases from 15.1 to 13.6 (see Fig. 1c).

The *ab initio* phasing was carried out on each dataset by the charge-flip algorithm using the SUPERFLIP program. Initial structure models were then derived from the resulting electron densities using the JANA2006 program, and both were found to be isostructural to Cd$_6$R (R = rare-earth element) 1/1 ACs, which is described as a body-centered cubic packing of Tsai-type RTH clusters [15]. The atomic coordinates, site occupation factors (SOFs) and equivalent isotropic ADPs ($U_{eq}$) are summarized in Tables 2 and 3 for the $y = 13.6$ and 15.1 samples, respectively. The cluster is comprising following successive shells; the second dodecahedron shell (M1 and M5), third icosahedron shell (Ce1), fourth icosidodecahedron shell (M3 and M4), and fifth RTH shell (M1, M2, M3 and M6). The M1 and M2 sites are shared by the second and fourth shells, respectively because two RTH clusters are interpenetrated with each other along a three-fold axis. The first shell was absent at this stage. Furthermore, an additional atom was

Table 1 Crystallographic information and parameters for the data collection and structural refinement of the $y = 13.6$ and 15.1 for (Au$_{0.70}$Ga$_{0.30}$)$_{100-y}$Ce$_y$ 1/1 approximants

| $y$ for (Au$_{0.70}$Ga$_{0.30}$)$_{100-y}$Ce$_y$ | 13.6 | 15.1 |
|---|---|---|
| Crystal Data | | |
| nominal composition | Au$_{107.90}$Ce$_{24}$Ga$_{44.41}$ | Au$_{105.35}$Ce$_{25.16}$Ga$_{42}$ |
| molar mass (g/mol) | 27712.02 | 27204.69 |
| temperature (K) | 298(2) | |
| space group | $Im\bar{3}$ | |
| $a$ (Å) | 14.88720(5) | 14.82600(3) |
| cell volume (Å$^3$) | 3299.43(3) | 3258.907(19) |
| crystal form | Prism | Prism |
| crystal size ($\mu$m) | 50*60*50 | 51*61*53 |
| $Z$ | 1 | |
| $F(000)$ | 11292.7 | 11084.0 |
| calculated density (g/cm$^3$) | 13.947 | 13.862 |
| Data Collection | | |
| radiation type | Mo K$\alpha$ | |
| $\theta_{min}$, $\theta_{max}$ (deg.) | 2.736, 46.455 | 1.943, 45.255 |
| no. of measured reflections | 48248 | 45550 |
| no. of independent reflections | 2631 | 2446 |
| no. of observed reflections | 2493 | 2384 |
| absorption coefficient (mm$^{-1}$) | 136.321 | 135.069 |
| Refinement | | |
| refined composition | Au$_{54.00(18)}$Ce$_{12.04}$Ga$_{21.90(26)}$ | Au$_{61.07(13)}$Ce$_{14.58}$Ga$_{24.34(9)}$ |
| $R_{int}$ | 0.0652 | 0.0871 |
| no. of parameters | 75 | 61 |
| $R_1$ | 0.0295 | 0.0286 |
| w$R_2$ | 0.0589 | 0.0730 |
| $S$ | 1.140 | 1.156 |

…

Table 2 The atomic coordinates, SOFs and equivalent isotropic displacement parameters $U_{eq}$ for $y = 13.6$ of $(Au_{0.70}Ga_{0.30})_{100-y}Ce_y$

| Site | Element | Wyckoff position | SOF | $x$ | $y$ | $z$ | $U_{eq}$ (Å$^2$) |
|---|---|---|---|---|---|---|---|
| Ce1 | Ce | 24g | 1 | 0.00000(2) | 0.18780(2) | 0.30521 | 0.004587 |
| M1 | Au | 16f | 1 | 0.14998(2) | 0.14998(2) | 0.14998(2) | 0.01184 |
| M2 | Au | 24g | 1 | 0.00000(2) | 0.40415 | 0.35719(2) | 0.006207 |
| M3 | Au | 48h | 1 | 0.10411(2) | 0.34198(2) | 0.20020(2) | 0.009217 |
| M4A | Au/Ga | 12d | 0.3748/0.3947 | 0.39751 | 0.00000(8) | 0 | 0.019433 |
| M4B | Au/Ga | 12d | 0.0963/0.1014 | 0.427 | 0.0000(4) | 0 | 0.0109 |
| M4C | Au/Ga | 24g | 0.0159/0.0168 | 0.0000(8) | 0.4265 | 0.0209(9) | 0.0011 |
| M5A | Au/Ga | 24g | 0.0717/0.1107 | 0.0000(4) | 0.2344(5) | 0.0785 | 0.0053 |
| M5B | Au/Ga | 24g | 0.1367/0.2111 | 0.00000(13) | 0.26098(14) | 0.07153 | 0.0053 |
| M5C | Au/Ga | 24g | 0.1846/0.2851 | 0.00000(16) | 0.2338(4) | 0.09198 | 0.024833 |
| M6 | Ga | 12e | 1 | 0.19999(7) | 0 | 0.5 | 0.007933 |
| M7A | Au/Ga | 24g | 0.0804/0.0647 | 0.0000(4) | 0.0884(4) | 0.0842 | 0.0269 |
| M7B | Au/Ga | 24g | 0.1043/0.0839 | 0.0000(4) | 0.0400(4) | 0.091 | 0.048333 |
| M8 | Ga | 8c | 0.99 | 0.25 | 0.25 | 0.25 | 0.0162 |

Table 3 The atomic coordinates, SOFs and equivalent isotropic displacement parameters $U_{eq}$ for $y = 15.1$ of $(Au_{0.70}Ga_{0.30})_{100-y}Ce_y$

| Site | Element | Wyckoff position | SOF | $x$ | $y$ | $z$ | $U_{eq}$ (Å$^2$) |
|---|---|---|---|---|---|---|---|
| Ce1 | Ce | 24g | 1 | 0.00000(2) | 0.18756(2) | 0.30464 | 0.004317 |
| Ce2 | Ce | 2a | 0.581 | 0 | 0 | 0 | 0.021 |
| M1 | Au | 16f | 1 | 0.14638(2) | 0.14638(2) | 0.14638(2) | 0.01601 |
| M2 | Au | 24g | 1 | 0.00000(2) | 0.40418(2) | 0.35658 | 0.00616 |
| M3 | Au | 48h | 1 | 0.10358(2) | 0.34088(2) | 0.19884(2) | 0.009683 |
| M4A | Au/Ga | 12d | 0.425/0.365 | 0.39667 | 0 | 0.00000(6) | 0.0156 |
| M4B | Au/Ga | 12d | 0.1127/0.0969 | 0.4248 | 0 | 0.0000(4) | 0.0156 |
| M5A | Au/Ga | 24g | 0.199/0.434 | 0.00000(10) | 0.22917 | 0.08706(10) | 0.014767 |
| M5B | Au/Ga | 24g | 0.1153/0.2516 | 0.00000(16) | 0.25587 | 0.0751(2) | 0.015 |
| M6 | Ga | 12e | 1 | 0.19916(7) | 0 | 0.5 | 0.007667 |
| M7A | Au | 24g | 0.0629 | 0.0000(13) | 0.0942 | 0.0902(13) | 0.067 |
| M7B | Au | 24g | 0.0769 | 0.0000(10) | 0.0429 | 0.0844(11) | 0.067 |
| M8 | Ga | 8c | 1 | 0.25 | 0.25 | 0.25 | 0.0195 |

observed at (1/4,1/4,1/4) (8c) for both samples (hereafter, labelled as M8). Such an additional site has been seen in Cd$_6$Ce 1/1 AC [45]. Because the additional site in the Cd$_6$Ce was observed to be occupied by a smaller atom (i.e. Cd), the mixture of Al/Ga was assigned to the M8 in the present Au-Ga-Ce 1/1 ACs.

The atomic coordinates, SOFs, and atomic displacement parameters (ADPs) were refined against 2493 and 2384 independent reflections [$F_{obs} \geq 2\sigma(F_{obs})$] for the $y = 13.6$ and 15.1 samples, respectively, utilizing the SHELXL program [46]. While the refinement cycles, the occupational and positional disorders were carefully analysed by introducing chemical mixing and site splitting.

After the iteration of the refinement cycles for the initial models, the difference Fourier synthesis yielded residual electron density at

…

M7 (24g) for the $y = 13.6$ sample. The SOF for M7 site (24g) was fixed to be 1/3 in a further refinement so that the number of atoms at the innermost shell is four. For the $y = 15.1$ sample, the difference Fourier synthesis yielded large residual electron density at (0,0,0) (2a) and weak but significant residual densities at M7 site. In a further refinement, Ce atom was assigned to the (0,0,0) position (hereafter, labelled as Ce2), and the SOFs of Ce2 and M7 were refined under a constrain so as a first shell being either a single Ce atom or a tetrahedron shell. In addition, the difference Fourier synthesis yielded residual electron densities around M4, M5 and M7 sites, therefore, these sites were splitted into two or more in the further refinement.

The subsequent refinements converged with reliability indices of ($R1$, $wR2$, $S$) = (0.0295, 0.0589, 1.140) and (0.0286, 0.0730, 1.156)

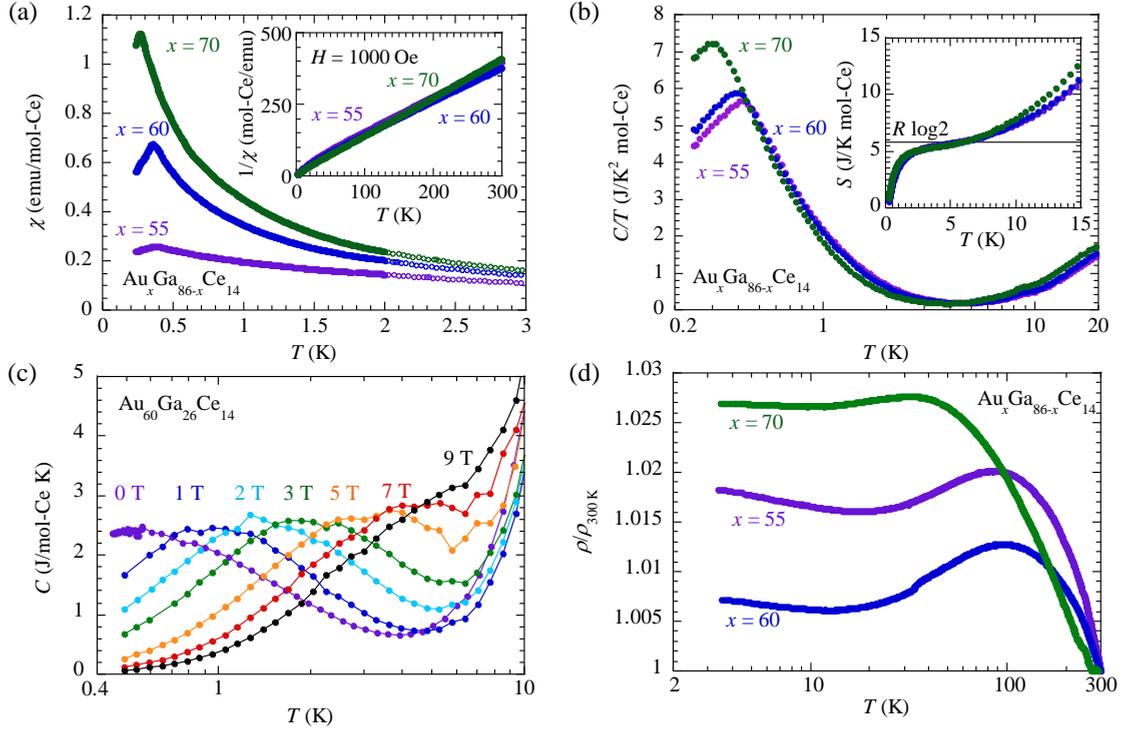

Figure 3 (a) The temperature dependence of magnetic susceptibility $\chi$ on Au$_x$Ga$_{86-x}$Ce$_{14}$. All measurement is performed with zero field cooling. AC and DC measurement results are represented by filled and open circle, respectively. Insets show the inverse magnetic susceptibility $1/\chi$ up to 300 K for various $x$. (b) The temperature dependence of the specific heat ($C$) divided by the temperature ($C/T$) of Au$_x$Ga$_{86-x}$Ce$_{14}$. Inset shows temperature dependence of the entropy estimated by $C/T$ as $S = \int \frac{C}{T} dT$ with assuming $S(T=0) = 0$. The black line refers $R \log 2 = 5.8$ J/K mol-Ce value. (c) The temperature of heat capacity $C$ of Au$_{60}$Ga$_{26}$Ce$_{14}$ under various magnetic field. (d) The temperature dependence of resistivity $\rho$ of Au$_x$Ga$_{86-x}$Ce$_{14}$. All $\rho$ are normalized by dividing by the value of 300 K.

for the $y = 13.6$ and 15.1 samples, respectively. Figures 2 shows the shell structures of the Tsai-type RTH cluster for the $y = 13.6$ and 15.1 samples. As seen from the figure, preferential site occupation of the constituent elements is noticed for both the structures: First, the M2 site at the second dodecahedron shell, M1 site at the fourth icosidodecahedron shell and M3 site at the fifth RTH shell are fully occupied by Au; Second, the M6 site at the fifth RTH shell and M8 site are occupied by Ga; Third, the Ce1 site at the third icosahedron shell is occupied by Ce. On the other hand, the structures exhibit occupational and positional disorders at M4 and M5 sites, both of which are occupied by non-magnetic Au/Ga mixtures. For the sample with $y = 13.6$, the M4 and M5 is occupied by Au/Ga with mixing ratios of 0.50/0.50 and 0.40/0.60, respectively. The former splits into M4a, M4b and M4c, separated by 0.62(3) Å and the later splits into M5a and M5b, separated by 0.506(5) Å. For the sample with $y = 15.1$, the M4 and M5 is occupied by Au/Ga with mixing ratios of 0.54/0.46 and 0.69/0.31, respectively. The former splits into M4a and M4b, separated by 0.417(5) Å and the later splits into M5a and M5b, separated by 0.434(4) Å.

The atomic structures of the $y = 13.6$ and 15.1 samples are distinguishable from each other by the presence or absence of the magnetic Ce atom at the single Ce2 site. For the $y = 13.6$ sample, no residual peak was observed (0,0,0) on the difference Fourier map indicating the absence of the Ce atom at Ce2. Instead, M7 site is partially occupied by Au/Ga with a mixing ratio of 0.55/0.45. The M7 site splits into M7a and M7b, and the total SOFs of these split sites equals to 1/3, meaning that the four atoms exist at the innermost shell. For the sample with $y = 15.1$, the single site Ce2 is partially occupied by Ce with a SOF being 0.581(3), which indicates that approximately 58% of clusters have the magnetic Ce atom at Ce2 site instead of the non-magnetic tetrahedron shell at M7. The M7 site splits into M7a and M7b, and each of which is occupied by pure Au, and total SOF of these split sites equals to 0.140(1). The decrease of the unit-cell parameter with the increase of Ce concentration in Fig. 1(c) is, therefore, explained by the increase of the SOF at Ce2 site by Ce atom whose size is smaller than that of the Au/Ga tetrahedron. Here, we note that the Ce occupation at Ce2 results not only in the decrease of the unit-cell parameter but also in an injection of an impurity Ce spin if its valence is not 4+.

In the Zn$_6$Sc 1/1 AC, which is isostructural to the present Al-Ga-Ce 1/1 ACs, it has been shown that the orientation of the tetrahedron shell is dynamically disordered at room temperature and invokes large distortion of the dodecahedron and icosidodecahedron shells [47,48]. Although, it is not clear whether the orientation of the non-magnetic Au/Ga tetrahedron shells is a dynamically or statically disordered in the present 1/1 ACs at present, the large splitting of M4 and M5 sites is attributed to large deviations from their average positions by the shell distortion.

### C. Physical properties

Figure 3(a) displays the temperature dependence of magnetic susceptibility $\chi$ for various values of $x$ in Au$_x$Ga$_{86-x}$Ce$_{14}$. Sharp cusps, which are a characteristic sign of spin freezing, are clearly observed at $T_f = 0.39$, 0.36, and 0.27 K for $x = 55$, 60, and 70, respectively. The inset of Fig. 3(a) shows the temperature dependence of inverse dc magnetic susceptibility $1/\chi$ for various $x$ between 2 and 300 K, where $T$-linear dependence of $1/\chi$ is observed above 100 K, showing a well-localized nature of Ce spins. Figure 3(b) shows the

…

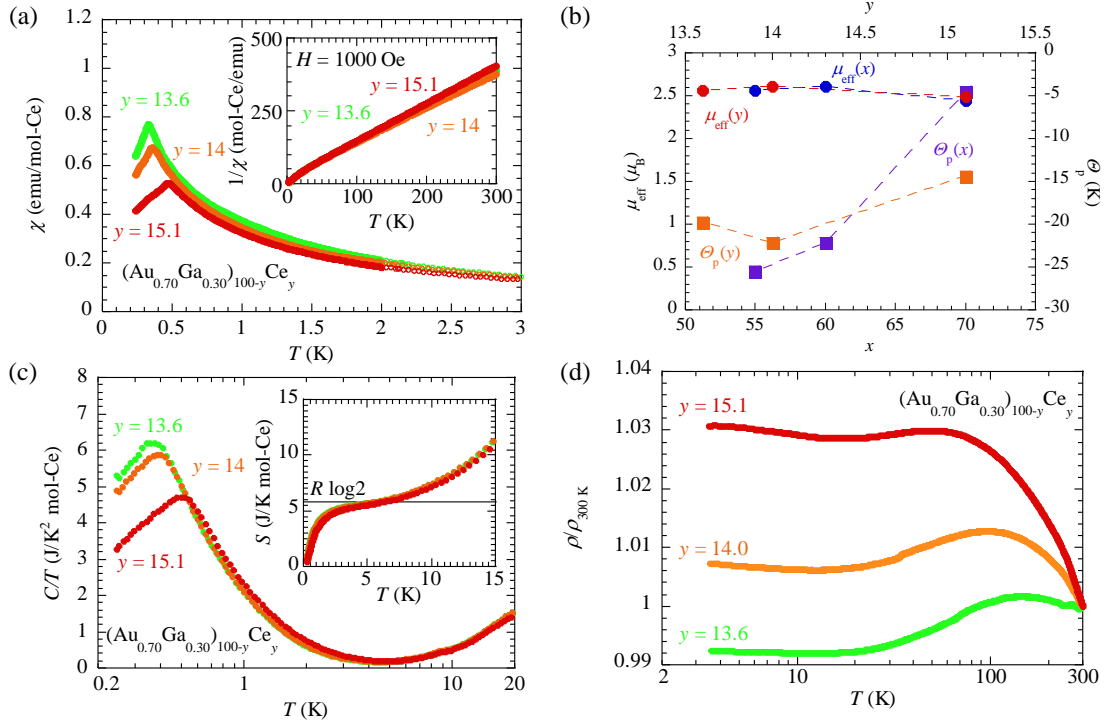

Figure 4 (a) The temperature dependence of magnetic susceptibility $\chi$ on $(Au_{0.70}Ga_{0.30})_{100-y}Ce_y$. All measurement is performed with zero field cooling. AC and DC measurement results are represented by filled and open circle, respectively. Insets show the inverse magnetic susceptibility $1/\chi$ up to 300 K for various $x$. (b) The $x$ and $y$ dependence of effective moment $\mu_{eff}$ and paramagnetic Curie temperature $\theta_p$ estimated by Curie-Weiss fitting. (c)The temperature dependence of the $C/T$ of $(Au_{0.70}Ga_{0.30})_{100-y}Ce_y$. Inset shows temperature dependence of the entropy estimated by $C/T$ as $S = \int \frac{C}{T}dT$ with assuming $S(T = 0) = 0$. The black line refers $R\log2 = 5.8$ J/K mol-Ce value. (d) The temperature dependence of $\rho$ of $(Au_{0.70}Ga_{0.30})_{100-y}Ce_y$. All $\rho$ are normalized by dividing by the value of 300 K.

temperature dependence of heat capacity ($C/T$) for various values of $x$ in $Au_xGa_{86-x}Ce_{14}$. A significantly large peak is commonly observed at 0.40(2), 0.38(2), and 0.30(2) K for $x$ = 55, 60, and 70, respectively, the temperatures of which are in good agreement with the $T_f$ observed in magnetization measurement. This implies that the large peak in $C/T$ is correlated with the magnetic anomaly. On the other hand, almost no dependence on $x$ is noticed in the $C/T$ behaviour above the cusp temperatures. We note that the temperature dependence of $C/T$ of the present Au-Ga-Ce 1/1 AC is identical to those of the Ag-In-Ce [40] and Au-Al-Ce [42] 1/1 ACs, which, on the other hand, indicates that such a gigantic $C/T$ is a common property of the bcc array of the Ce-icosahedron. The inset of Fig. 3(b) shows the entropy estimated from $S = \int \frac{C}{T}dT$ by linear extrapolation of $C/T$ below 0.3 K, i.e., the lowest temperature, toward zero at 0 K. A plateau with $\sim R\log2$ = 5.8 J/K mol-Ce is observed for all compounds, which indicates the ground state is doublet for all the samples. We note that the temperature region exhibiting the plateau shows almost no dependence on $x$. Fig. 3(c) displays the temperature dependence of heat capacity $C$ for $Au_{60}Ga_{26}Ce_{14}$ under magnetic field. The peak temperature clearly increases with increasing magnetic field, indicating the existence of strong ferromagnetic correlation in all the samples. Since a ferromagnetic transition is not observed in the $\chi$-$T$ curves, the magnetic ground state of the Au-Ga-Ce 1/1 AC is concluded to be a spin-glass state for all the studied values of $x$. Figure 3(d) shows the temperature dependence of the resistivity $\rho$ of the Au-Ga-Ce 1/1 AC, where the occurrences of a maximum and a minimum in $\rho$ with decreasing temperature is commonly observed. Such a maximum in $\rho$ is a characteristic feature of the formation of a Kondo lattice [49, 50]. Concerning the $\rho$ minimum, a log $T$ increase in $\rho$ is observed below the minimum temperature, the origin of which will be addressed later.

The temperature dependence of the magnetic susceptibility $\chi$ for various $y$ in $(Au_{0.70}Ga_{0.30})_{100-y}Ce_y$ is displayed in Fig. 4(a). Cusps are clearly observed at $T_f$ = 0.33, 0.36, and 0.47 K for $y$ = 13.6, 14.0, and 15.1, respectively, which indicates that the magnetic ground state remains spin glass for all the samples with different $y$ values, and the replacement of the tetrahedron by an Ce atom stabilizes the spin-glass state as inferred from the increase in the $T_f$, which is in agreement with the observation in the Au-Si-Tb 1/1 AC [30]. We also notice a decrease of the magnitude of $\chi$ with increasing the Ce concentration, which suggests that the Ce atom at the center of the cluster is screened by Kondo effect.

The inset of Fig. 4(a) shows the temperature dependence of $1/\chi$, where a $T$-linear dependence is clearly noticed above 100 K, also in agreement with the inset of fig. 3(a). Therefore, we performed the Curie-Weiss fitting with

$$\chi = \frac{N_A \mu_{eff}^2 \mu_B^2}{3k_B(T - \theta_p)}$$

, where $k_B$, $\theta_p$, $N_A$, $\mu_{eff}$ and $\mu_B$ are the Boltzmann constant, paramagnetic Curie temperature, Avogadro's number, effective moment, and Bohr magneton, respectively. The estimated $\mu_{eff}$ and $\theta_p$ are displayed in Fig. 4(b). The obtained values of $\mu_{eff}$ are ~2.54 $\mu_B$, independent of the chemical substitution and is close to the theoretical value of the $Ce^{3+}$ free ion, which indicates that the 4$f$ electrons are well localized at high temperatures and are not much

…

affected by the structural modifications. On the other hand, $\theta_p$ is negative over the large $e/a$ region and increases with Au substitution showing an increase of ferromagnetic correlation between Ce spins.

Figure 4(c) shows the temperature dependence of $C/T$ for various $y$ in $(Au_{0.70}Ga_{0.30})_{100-y}Ce_y$. A clear anomaly is observed at 0.36(2), 0.38(2), and 0.50(2) K for $y$ = 13.6, 14.0, and 15.1, respectively. Since these temperatures are close to the cusp temperatures of $\chi$, it also indicates that the anomalies are due to the spin-glass transition and, thus, the gigantic $C/T$ is due to spin-glass freezing. In addition, the magnitude of $C/T$ is found to decrease slightly with the Ce substitution, which suggests that the contribution of Ce at the cluster center to $C/T$ is lower implying that the $bcc$ lattice of the Ce icosahedron is mainly responsible for the such gigantic $C/T$ anomaly. The entropy $S$ estimated from the $C/T$ data using the same method used for obtaining the inset of fig 3(b) is shown in the inset of fig 4(c). A plateau of ~ $R \log 2$ is again observed, implying the formation of doublet. Note that the introduction of Ce into the cluster center has almost no contribution to the entropy behavior.

Figure 4(d) displays the temperature dependence of $\rho$ of the Au-Ga-Ce 1/1 ACs, where all the curves commonly exhibit a maximum, due to a formation of a Kondo lattice, and a minimum with decreasing temperature. Interestingly, the $\rho$ - $T$ behavior below the minimum is found to be strongly dependent on the value of $y$; a log $T$ dependence becomes more pronounced with increasing the value of $y$. There is almost no temperature dependence for the lowest value of $y$ = 13.6, where no Ce atom exists at the cluster center. Thus, such a log $T$ dependence of $\rho$ is brought about by a Ce atom introduced at the cluster center, implying that this Kondo effect is due to an impurity Ce spin at the cluster center.

### D. Discussion

The number of electrons per atom, $e/a$, has been reported to be an important parameter of the magnetism of Tsai-type 1/1 ACs since the magnetic ground state and $\theta_p$ can be well controlled by $e/a$ in the Gd and Tb ACs [21]. In the Au-Ga-Ce ACs, $e/a$ is found to be largely tunable from 1.60 to 1.94 by changing the Au concentration, where we assume that Ce is in the trivalent state ($Ce^{3+}$) as justified by the result of the Curie-Weiss fitting. First, the observation of spin-glass ground state with negative $\theta_p$ in the Ce based AC is consistent with the observations in Gd and Tb based ACs. One significant difference between Ce and Tb/Gd based ACs exists in the fact that spin glass ground state prevails over a wider $e/a$ region of $e/a$ = 1.60 – 1.94 for the Ce based AC while magnetically-ordered ground states appear when $e/a <$ 1.84 in Gd/Tb based ACs [21], the clarification of which has now become a future issue.

Concerning the possible origin of the spin-glass behavior of the Au-Ga-Ce 1/1 AC, both antiferromagnetic interaction and disorder are necessary for spin-freezing phenomena. For the Au-Ga-Ce 1/1 AC, negative $\theta_p$ over the entire single-phase region means the existence of considerable antiferromagnetic contributions for all the compositions. In addition, there exists intrinsic chemical disorder at non-magnetic atom sites, so-called non-magnetic atom disorder (NMAD), at the M4, M5, and M7 sites, as evidenced from the structural analyses summarized in Tables 2 and 3. Thus, the spin-glass behavior in the present Au-Ga-Ce 1/1 AC is understood to be triggered by NMAD.

For the gigantic $C/T$ peak commonly observed in the Au-Ga-Ce 1/1 ACs, it can be reasonably understood as a result of a Schottky peak, not as a heavy fermion behavior. Such a Schottky peak is often observed at low temperatures in small-moment systems such as Ce-based ones [51]. In order to further clarify this issue, investigations of low temperature physical properties of ACs without chemical disorder, e.g., $Cd_6Ce$, will be certainly of importance.

The partial Ce injection into the cluster center is found to enhance the log $T$ dependence of the low temperature resistivity, which implies that a Ce atom at the cluster center behaves as a Kondo impurity in the $bcc$ lattice formed by the Ce-icosahedron. To the authors' best knowledge, substitution of different rare-earth elements to Kondo lattice sites was performed in many compounds [52-54], however, one significant difference from the previous works exists in the fact that in the present work impurity rare-earth atoms are introduced not to the Kondo lattice sites but to non-magnetic atom sites. In this view, Tsai-type ACs provide a new playground for investigations of magnetic impurity effect in a Kondo lattice without disturbing the Kondo lattice and with precisely controlling the amount of magnetic impurity at a certain site.

### IV. SUMMARY

A new Ce-based Tsai-Type 1/1 AC was discovered in the Au-Ga-Ce ternary system. This 1/1 AC possesses a wide single-phase region in both Au/Ga and (Au,Ga)/Ce parameter spaces. Among the two degrees of structural freedom, the latter is related to the Ce occupancy at the center of the Tsai-type cluster as evidenced by the structural analyses. Magnetic susceptibility, heat capacity and resistivity measurements in the series of $Au_xGa_{86-x}Ce_{14}$ and $(Au_{0.70}Ga_{0.30})_{100-y}Ce_y$ show that the magnetic ground state is a spin-glass state, not being affected by the substantial variations in both parameters. The occurrence of the spin-glass state is understood in terms of the combined effect of dominant antiferromagnetic interaction, i.e., the negative paramagnetic Curie temperature, and non-magnetic atom disordering as evidenced by the structural analyses. In this view, the extraordinarily large $C/T$ observed in the Ce-based AC is attributed not to a heavy-Fermion nature but to spin-freezing phenomenon. Furthermore, a spin injection into the cluster center is found to give rise to the spin-glass formation and Kondo effect, the latter of which indicates that an injected Ce spin behaves as a Kondo impurity in the magnetic $bcc$ lattice formed by the Ce icosahedron.

### ACKNOWLEDGMENT

We would like to thank Y. Muro, T. Sugimoto, G. Nozue and A. Sekiyama for fruitful discussions. This work was supported by JSPS KAKENHI (Grants No. JP19K14663, No. JP19H05817, No. JP19H05818, No. JP19H05821) and JST, CREST Grant No. JPMJCR22O3, Japan. This work was partly carried out under the Visiting Researcher's Program of the Institute for Solid State Physics, the University of Tokyo and Cooperative Research Program of "Network Joint Research Center for Materials and Devices".

…

…